\documentclass[prl,twocolumn,10pt,superscriptaddress]{revtex4-2}

\usepackage{amsmath}
\usepackage{amssymb}
\usepackage{mathtools}
\usepackage{graphicx}
\usepackage[usenames,dvipsnames]{color}
\usepackage{bm}
\usepackage{hyperref}
\usepackage{url}
\usepackage[utf8]{inputenc}
\usepackage{subfigure}
\usepackage{slashed,bbm}
\usepackage{graphics,psfrag,epsfig}
\usepackage{dsfont}
\usepackage{setspace}
\usepackage{wasysym}

\usepackage{hyperref}
\usepackage{xcolor}
\definecolor{dark-red}{rgb}{0.4,0.15,0.15}
\definecolor{dark-blue}{rgb}{0.15,0.15,0.4}
\definecolor{medium-blue}{rgb}{0,0,0.5}
\hypersetup{
colorlinks, linkcolor={dark-blue},
citecolor={dark-blue}, urlcolor={medium-blue}
}

\newcommand{\Nf}{N_\mathrm{f}}

\newcommand{\hc}{\text{h.c.}}

\newcommand{\ii}{\text{i}}
\renewcommand{\vec}[1]{\mathbf{#1}}
\newcommand{\p}{\text{p}}
\DeclareMathOperator{\tr}{tr}

\usepackage[normalem]{ulem}

\begin{document}

\title{Projection of Infinite-$U$ Hubbard Model and Algebraic Sign Structure}

\author{Yunqing Ouyang}
\affiliation{State Key Laboratory of Surface Physics, Fudan University, Shanghai 200433, China}
\affiliation{Center for Field Theory and Particle Physics, Department of Physics, Fudan University, Shanghai 200433, China}
\author{Xiao Yan Xu}
\email{xiaoyanxu@sjtu.edu.cn}
\affiliation{Key Laboratory of Artificial Structures and Quantum Control (Ministry of Education), School of Physics and Astronomy, Shanghai Jiao Tong University, Shanghai 200240, China}
\affiliation{Department of Physics, University of California at San Diego, La Jolla, California 92093, USA}

\date{\today}

\begin{abstract}
We propose a projection approach to perform quantum Monte Carlo (QMC) simulation on the infinite-$U$ Hubbard model at some integer fillings where either it is sign problem free or surprisingly has an algebraic sign structure --  a power law dependence of average sign on system size. We demonstrate our scheme on the infinite-$U$ $SU(2N)$ fermionic Hubbard model on both a square and honeycomb lattice at half-filling, where it is sign problem free, and suggest possible correlated ground states. The method can be generalized to study certain extended Hubbard models applying to cluster Mott insulators or two-dimensional Moir\'e systems; among one of them at certain non-half-integer filling, the sign has an algebraic behavior such that it can be numerically solved within a polynomial time. Further, our projection scheme can also be generalized to implement the Gutzwiller projection to spin basis such that $SU(2N)$ quantum spin models and Kondo lattice models may be studied in the framework of fermionic QMC simulations.
\end{abstract}

\maketitle
{\it Introduction}\,---\,
The Hubbard model~\cite{Gutzwiller1963effect,Hubbard1963electron,Arovas2021hubbard,Qin2021hubbard} provides a starting point to understand physics in strongly correlated electronic systems, such as cuprates~\cite{Anderson1987the,Emery1987theory,Dagotto1994corr,
Imada1998metal,Lee2006doping}, iron-based superconductors~\cite{Hirschfeld2011gap,Fernandes2016low}, and heavy-fermion materials~\cite{Schrieffer1966relation,Scalapino2012a}, as well as ultracold-atom-simulated correlated systems~\cite{Mazurenko2017a} and recently found multiple Mori\'e superlattice systems~\cite{
Balents2020superconductivity,Andrei2021the,Li2021continuous}. In those strongly correlated electronic systems, the possible phases of the strong correlation limit are extremely important, outlining possible topology of phase diagrams or serving as mother states to generate more exotic phases.

The infinite-$U$ Hubbard model provides an important perspective on strong correlation physics~\cite{Nagaoka1966ferr,Liu2012phases,Tandon1999comp}. Solid conclusions can only be made on limited and specific cases, e.g., Nagaoka's theorem~\cite{Nagaoka1966ferr} applies to low hole density limit of the infinite-$U$ Hubbard model on bipartite lattice and Lieb's theorem~\cite{Lieb1989two} imposes constraints on the ground state of the Hubbard model with bipartite hopping at half filling. While numerical methods may provide important hints~\cite{Liu2012phases,Tandon1999comp}, disputes still exist on questions such as which phase is the true ground state of the large-$U$ $SU(N)$ Hubbard model on several lattices~\cite{Assaad2005phase,Paramekanti2007sun,Zhou2016mott,Zhou2018mott,Kim2019dimensional}.
Recently, large-$U$ $SU(N)$ Hubbard models are getting more and more attention in ultracold atom simulations~\cite{Taie2012an,Cai2013pomeranchuk,
Hofrichter2016direct,Ozawa2018Antiferromagnetic,
Taie2020observation,Eduardo2021universal,
Tusi2021flavour,Altman2021quantum}. 

Alternatively, an infinite-$U$ Hubbard model can be used as a constraint on local Hilbert space; a typical example is a study of Kondo lattice models, where the local spin is written in terms of fermion operators and a constraint is imposed to restore the local spin Hilbert space by introducing a Hubbard-$U$ term~\cite{Assaad1999quantum,Raczkowski2020phase}. A finite Hubbard-$U$ term plays as a soft constraint, while if $U$ goes to infinity, it becomes an exact constraint, and an elegant form for the $SU(2)$ case is pointed out in Ref.~\cite{Capponi2001spin}.   

Inspired by the mentioned former works~\cite{Assaad1999quantum,Capponi2001spin}, in this Letter we propose a general projection approach, such that controllable large-scale quantum Monte Carlo (QMC) simulations on various infinite-$U$ Hubbard models may be done at some integer fillings. Our scheme can be well demonstrated on the infinite-$U$ $SU(2N)$ fermionic Hubbard models on both square and honeycomb lattices at integer fillings, and the Monte Carlo results are presented at half filling where it is sign problem free. We found the infinite-$U$ $SU(4)$ Hubbard model with Dirac dispersion on square and honeycomb lattices may stabilize a spin liquid (SL) state.  
We further show how to generalize our scheme to study extended Hubbard models, such as a cluster charge Hubbard model on both square and honeycomb lattices, and obtain possible ground states at half-filling. Finally, we apply the projection approach to more exotic $SU(4)$ extended Hubbard models with only an interaction term. It is sign problem free at half filling and it has an $SU(4)$ ferromagnetic ground state. For a certain non-half integer filling, there is a sign problem, but the average sign happens to be only power law dependence on system size, such that it is also simulatable. This finding inspires a different perspective on finding Monte Carlo simulatable models.

\begin{figure}[t]
\centering
\includegraphics[width=0.95\hsize]{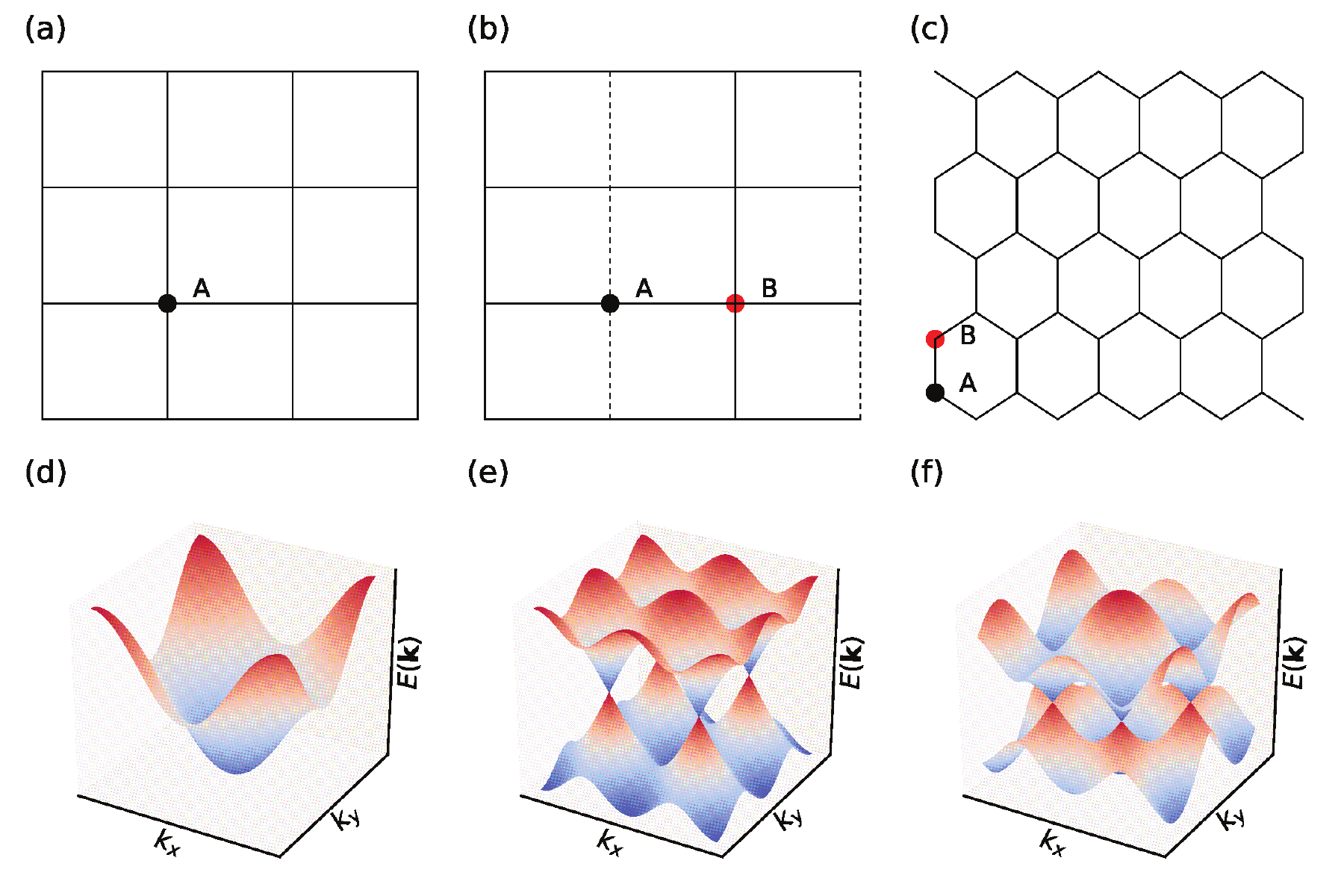}
\caption{Different fermiology considered. (a) Square lattice with uniform hopping. (b) Square lattice with $\pi$-flux hopping. We choose a gauge where the solid line denotes $t_{ij}=t$ and the dashed line denotes $t_{ij}=-t$.  (c) Honeycomb lattice with uniform hopping. (d) Energy band for (a) with nesting FS, denoted as $\square$-nesting-FS.  (e) Energy band for (b) with Dirac dispersion, denoted as $\square$-Dirac.  (f) Energy band for (c) with Dirac dispersion, denoted as $\varhexagon$-Dirac.}
\label{fig:fig1}
\end{figure}

{\it Projection Approach}\,---\,
We implement our projection approach in the framework of determinant QMC (DQMC)~\cite{Blankenbecler1981monte}. We illustrate our projection approach through an $SU(\Nf)$ Hubbard model with Hamiltonian $H=H_t+H_U$ on a general lattice, with the kinetic part $H_t=-\sum_{ij\alpha}[t_{ij}c_{i,\alpha}^\dagger c_{j,\alpha} + \text{H.c.}]$, and Hubbard interaction part $H_U=\frac{U}{2} \sum_{i}(n_{i}- \nu)^2$. Here the fermion density operator $n_i=\sum_{\alpha} n_{i,\alpha}$ at each site is a sum over fermion flavor density operator $n_{i,\alpha}=c_{i,\alpha}^\dagger c_{i,\alpha}$ with $\alpha=1,\cdots,\Nf$. We focus on the repulsive Hubbard interaction case($U>0$). In DQMC, one starts with partition function $Z=\tr[e^{-\beta H}]$ and observables $\langle O \rangle = \tr[Oe^{-\beta H}] / Z $, where $\beta$, the inverse temperature, is Trotter decomposed into $L_\tau$ slices, i.e., $\beta=L_\tau \Delta_\tau$. One needs to further make a Trotter decomposition, i.e., $e^{-\Delta_\tau H} \approx e^{-\frac 1 2 \Delta_\tau H_t} e^{-\Delta_\tau H_U} e^{-\frac 1 2 \Delta_\tau H_t}$. For the integer-filling infinite-$U$ case, $H_U$ plays the role of constraint on local Hilbert space, i.e., it defines a projection operator, and we observe the following \textit{exact} relation in the infinite-$U$ limit~\cite{suppl}:
\begin{equation}
\left.e^{-\frac{\Delta_\tau U}{2} (n_{i}- \nu)^2}\right|_{U\rightarrow+\infty} =  \frac{1}{M}\sum_{s_i=1}^M e^{\frac{\ii 2\pi s_i}{M} (n_{i}- \nu)},
\end{equation}
with $M=\frac{\Nf}{2}+|\tilde{\nu}|+1$, and an effective filling $\tilde{\nu}\equiv \nu-\frac{\Nf}{2}$ in reference to half filling. As we only focus on integer fillings, the effective filling $\tilde{\nu}$ takes values $\tilde{\nu}=-\frac{\Nf}{2},-\frac{\Nf}{2}+1,\cdots,\frac{\Nf}{2}$, where $\tilde{\nu}=0$ ($\nu=\frac{\Nf}{2}$) corresponds to half-filling. The projection is done by introducing a sum over auxiliary fields $s_i$. With the above projection operator, the trace over fermions can be easily performed~\cite{Blankenbecler1981monte,AssaadEvertz2008}, and the partition function $Z=\sum_c w_c$ and observables $\langle O \rangle = \frac{\sum_c O_c w_c}{\sum_c w_c}$ depend on auxiliary fields $c=\{s_{i,l}\}$ ($l$ is the time slice index), and the sampling over auxiliary fields $\{s_{i,l}\}$ can be done with Monte Carlo simulation. We note the above finite temperature DQMC scheme can be easily adapted to the zero-temperature projection DQMC~\cite{Sugiyama1986,Sorella1989a,Sorella1988numerical}. As the auxiliary fields here only take finite values, a local update with Metropolis algorithm is efficient. One caution here on the sign problem: for the Hubbard model with bipartite hopping at half filling, one can easily prove the sign problem free~\cite{Wu2005sufficient}, and numerical rigorous results can be obtained; for some special cases we find very likely ground state candidates at certain non-half-integer fillings even with the sign problem.

\begin{figure}[t]
\centering
\includegraphics[width=0.95\hsize]{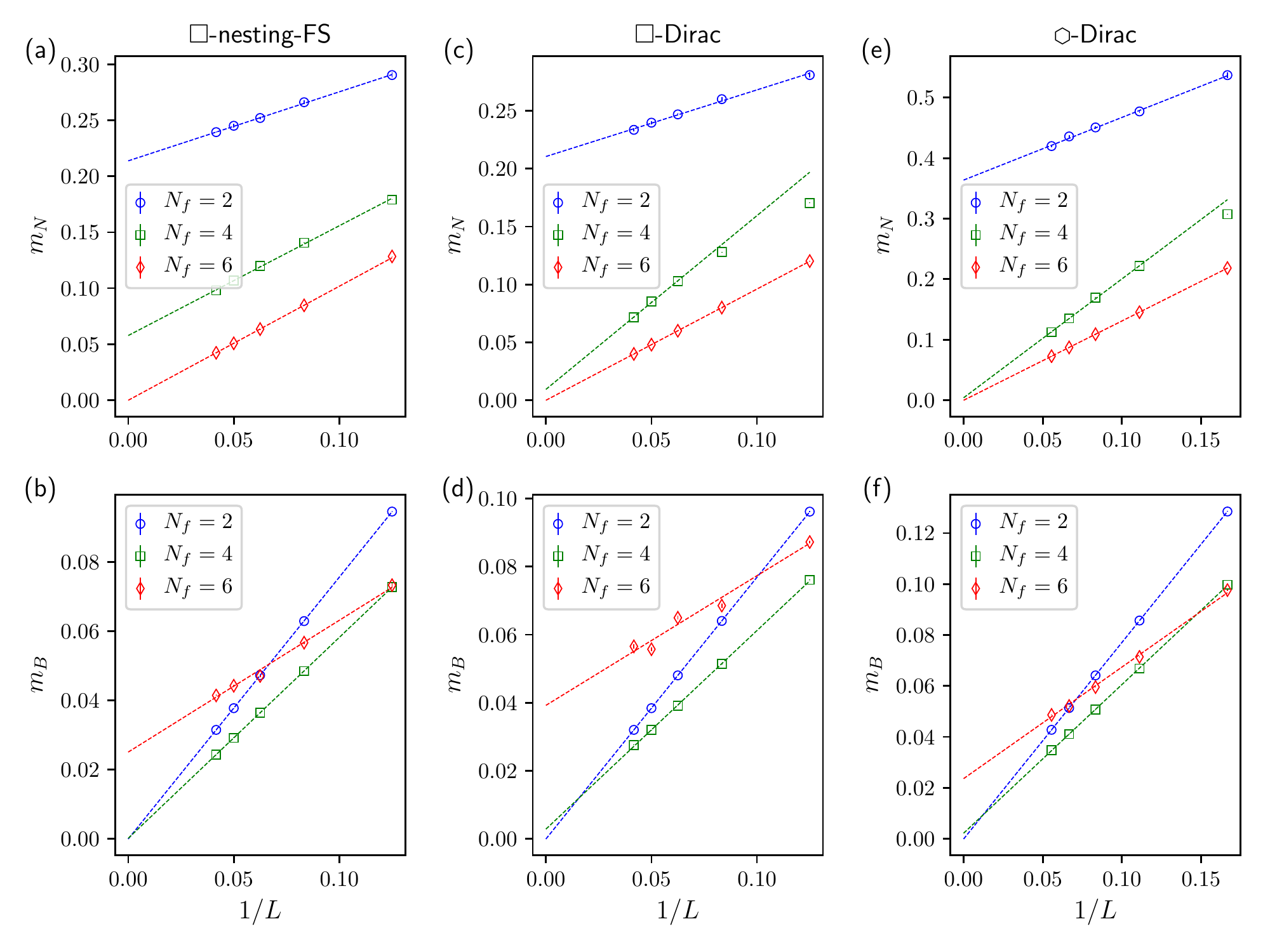}
\caption{$1/L$ extrapolation of N\'eel order parameter $m_N$ and VBS order parameter $m_B$. Panels (a) and (b) are for $\square$-nesting-FS, (c) and (d) for $\square$-Dirac, and (e) and (f) for $\varhexagon$-Dirac.}
\label{fig:fig2}
\end{figure}

{\it Infinite-$U$ $SU(2N)$ Hubbard model on bipartite lattice}\,---\,
We apply the above projection approach to the $SU(2N)$ Hubbard model both on two dimensional (2D) square and honeycomb lattices. We use a zero-temperature projection DQMC, with $L_\tau=600$ and $\Delta_\tau t=0.1$ in our calculation. We have performed about $0.3\times 10^4$ warmup sweeps and $1.2\times 10^4$ measurement sweeps (grouped into 20 bins) for each parameter running. For a square lattice, we consider two conventional fermionologies, one is the typical uniform nearest neighbor hopping which gives a nesting Fermi surface at half filling (denoted as $\square$-nesting-FS) as shown in Figs.~\ref{fig:fig1}(a) and (d), and the other is $\pi$-flux hopping, which gives Dirac dispersion (denoted as $\square$-Dirac) as shown in Figs.~\ref{fig:fig1}(b) and (e). For the honeycomb lattice, we consider uniform nearest neighbor hopping which also gives Dirac dispersion (denoted as $\varhexagon$-Dirac) as shown in Figs.~\ref{fig:fig1}(c) and (f).  We performed simulations on the infinite-$U$ $SU(2N)$ Hubbard model with the above mentioned fermionology, and identify a possible ground state as shown in Table~\ref{table:tab1}. 

\begin{table}[ht]
    \caption{Possible ground state of infinite-$U$ $SU(2N)$ Hubbard model at half filling on bipartite lattice with different fermiology. In the table, $U^\infty$ denotes infinite $U$.} 
    \centering 
    \def\arraystretch{1.5}
    \begin{tabular}{c  c  c  c} 
    \hline\hline 
     $SU(2N)$-$U^\infty$  & \ \  $\square$-nesting-FS \ \  & \ \ $\square$-Dirac \ \ & \ \ $\varhexagon$-Dirac \\ [0.0ex] 
    \hline 
     $SU(2)$   & \ \  N\'eel  &  \ N\'eel  & N\'eel  \\ [0.0ex] 
    \hline
     $SU(4)$  & \ \  N\'eel  &  \  SL?  & SL? \\ [0.0ex] 
    \hline
     $SU(\ge 6)$  & \ \  VBS  &  \  VBS  & VBS  \\ [0.0ex]
    \hline                    
    \end{tabular}
    \label{table:tab1} 
\end{table}

Before we discuss more details of all cases listed in Table~\ref{table:tab1}, we consider some analytical arguments. For the $SU(2N)$ Hubbard model on bipartite lattice, one can perform a $t/U$ expansion, giving Heisenberg interaction $J\sum_{\langle ij \rangle} \vec{S}_i \cdot \vec{S}_j$ at second order, with effective exchange coupling $J\sim \frac{t^2}{U}>0$. Therefore, the $SU(2N)$ Heisenberg model may capture some physics of the $SU(2N)$ Hubbard model at half filling, but be cautioned that they are different and may have different ground states in the infinite-$U$ limit, as the $t/U$ expansion will give zero $J$ in that limit. For the $SU(2)$ case, it is well known that the $SU(2)$ Heisenberg model on bipartite lattice has a N\'eel type ordered ground state. For $N$ larger than a certain value, it has a valence bond solid (VBS) order (also called spin-Peierls state)~\cite{Read1990spin,Harada2003neel,
Beach2009sun,Lang2013dimerized,Zhou2016mott,Li2017fermion}, and the critical $N_c$ is estimated about $2N_c=4.57(5)$ through QMC calculations~\cite{Beach2009sun}. Comparing with our numerics, the $SU(2)$ and $SU(\ge 6)$ are quite consistent with the $SU(2N)$ Heisenberg model, while the $SU(4)$ case is very special. For $\square$-nesting-FS, N\'eel type order is favored also for $SU(4)$, but for $\square$-Dirac and $\varhexagon$-Dirac, it is very likely that a SL state is stabilized. 

\begin{figure}[t]
\centering
\includegraphics[width=0.95\hsize]{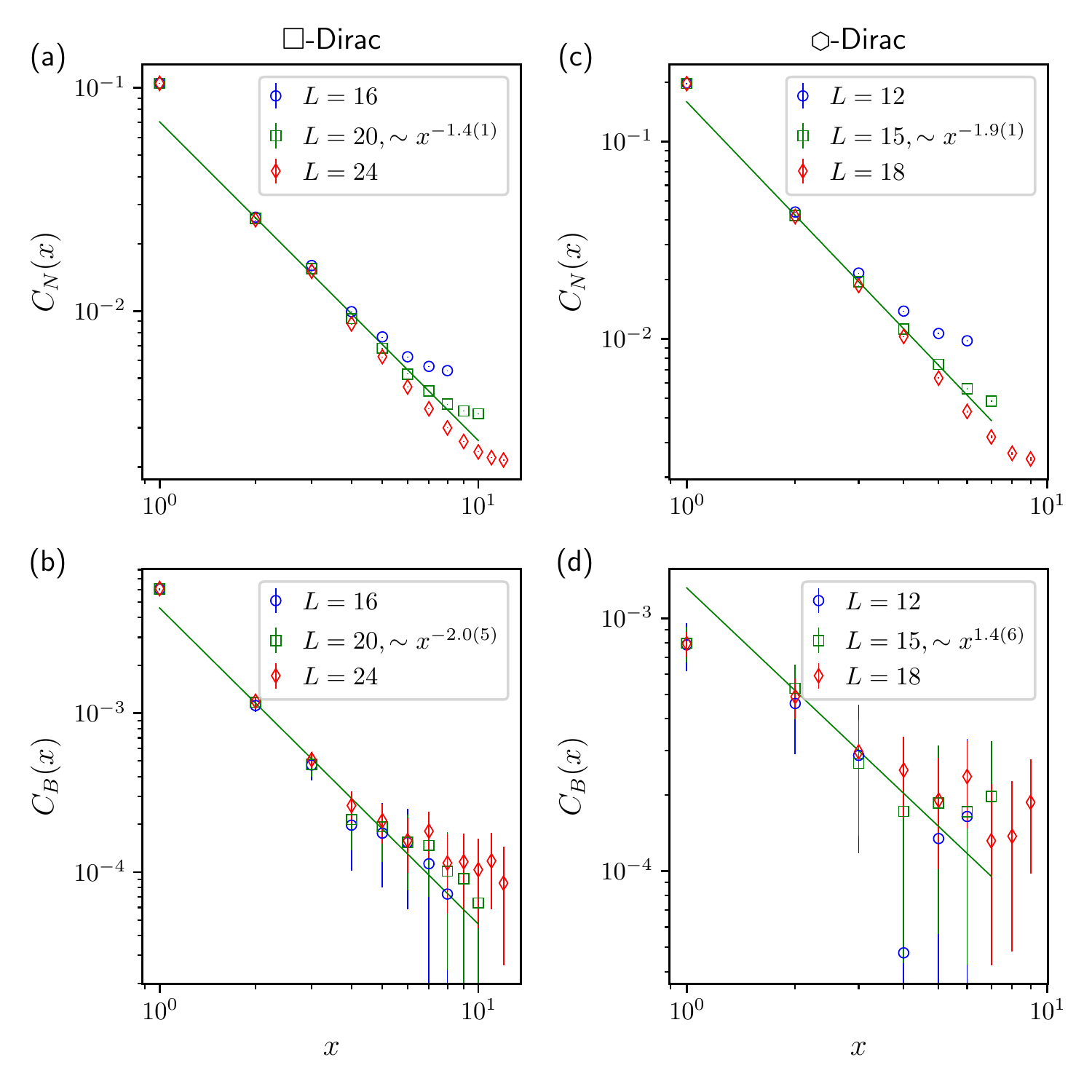}
\caption{Real space decay of N\'eel and VBS correlations. Panels (a) and (b) are for $\square$-Dirac and (c) and (d) for $\varhexagon$-Dirac. The green solid line is a power law fitting of $L=20$ data points ($x=2,\cdots,7$) for $\square$-Dirac and $L=15$ data points ($x=2,\cdots,5$) for $\varhexagon$-Dirac. Due to size limit, a systematic finite size scaling of the power law decay still cannot be realized.}
\label{fig:fig3}
\end{figure}

In the following, we investigate possible ordered states. One possible order is N\'eel type spin order. As the generators of $SU(\Nf)$ can be written as  $S_\beta^\alpha(\vec{r}_i)\equiv c_{i,\alpha}^\dagger c_{j,\beta} - \frac{\delta_{\alpha,\beta}}{\Nf}\sum_\gamma c_{i,\gamma}^\dagger c_{i,\gamma}$,
the matrix form of the N\'eel operator can be defined as $N_\beta^\alpha(\vec{r}_i) \equiv \frac{1}{\Nf} e^{-\ii \vec{Q} \cdot \vec{r}_i}S_\beta^\alpha(\vec{r}_i)$ with $\vec{Q}=(\pi,\pi)$ on square lattice, and $N_\beta^\alpha(\vec{r}_i) \equiv \frac{1}{\Nf} ( S_\beta^\alpha(\vec{r}_i+\boldsymbol{\tau}_1) - S_\beta^\alpha(\vec{r}_i+\boldsymbol{\tau}_2) )$ on honeycomb lattice, where $\boldsymbol{\tau}_1$ and $\boldsymbol{\tau}_2$ are inner-cell coordinates of two independent sites of each unit cell of honeycomb lattice. Another possible order is the VBS order, with the gauge invariant bond operator defined as $B(\vec{r}_i) \equiv \frac{1}{\Nf}e^{-\ii\vec{Q}\cdot\vec{r}_i}\sum_\alpha t_{i,i+\delta} c_{i,\alpha}^\dagger c_{i+\delta,\alpha}+\hc$, where $i,i+\delta$ are a pair of sites of two ends of nearest neighbor (NN) bond in a fixed direction, with $\vec{Q}=(\pi,0)$ corresponding to columnar VBS for square lattice, and $\vec{Q}=(\frac{2}{3}\pi,\frac{2}{3}\pi)$ for Kekul\'e type VBS for honeycomb lattice.
In the simulation, we measure N\'eel operator correlations $C_N(\vec{r}_i-\vec{r}_j)=\sum_{\alpha\beta}\langle N^\alpha_\beta(\vec{r}_i)  N^\beta_\alpha(\vec{r}_j) \rangle-\langle N^\alpha_\beta(\vec{r}_i)\rangle  \langle N^\beta_\alpha(\vec{r}_j) \rangle$ as well as bond operator correlations $C_B(\vec{r}_i-\vec{r}_j)=\langle B(\vec{r}_i) B(\vec{r}_j) \rangle - \langle B(\vec{r}_i) \rangle \langle B(\vec{r}_j) \rangle$.  With those correlations, we can extract N\'eel order parameter $m_{N}$ and the VBS order parameter $m_{B}$. The square of the N\'eel order parameter can be calculated as $m_N^2 = \frac{1}{L^4}\sum_{i,j} C_N(\vec{r}_i-\vec{r}_j)$, and the square of the VBS bond order parameter can be calculated as $m_B^2=\frac{1}{L^4}\sum_{i,j} C_B(\vec{r}_i-\vec{r}_j)$.  As shown in Fig.~\ref{fig:fig2}, we plot the $1/L$ extrapolation of the N\'eel and VBS order parameters for $SU(2)$, $SU(4)$,  and $SU(6)$ infinite-$U$ Hubbard models with different fermiology. For $SU(2)$, we have a finite N\'eel order parameter and, for $SU(6)$, we have a finite VBS order parameter, while, for $SU(4)$, we have a finite N\'eel order parameter for $\square$-nesting-FS, but for $\square$-Dirac and $\varhexagon$-Dirac both the N\'eel and VBS order are very likely zero in the thermodynamic limit. The softening of static form factors for both spin and bond further rules out any possible trend to ordinary magnetic orders for the infinite-$U$ $SU(4)$ Hubbard model with Dirac dispersion. We conjecture a SL may be stablized here. To further identify the possible ground state of the infinite-$U$ $SU(4)$ Hubbard model with Dirac dispersion, we plot the real space decay of spin-spin and bond-bond correlations as shown in Fig.~\ref{fig:fig3}, and found they have algebraic behavior approximately, which may indicate an algebraic SL state~\cite{Lee2006doping,Hermele2004stability,Hermele2005algebraic,Xu2019monte,Song2020from,Calvera2021theory}. Based on limited system sizes, it is hard to determine precisely the scaling dimensions of spin and bond operators and thus it is hard to tell whether the SL stabilized here is a Dirac $SU(4)$, $U(1)$, $Z_4$, or $Z_2$ SL~\cite{Lee2006doping,Hermele2004stability,Hermele2005algebraic,Xu2019monte,Song2020from,Calvera2021theory}.  The absence of a N\'eel or VBS order parameter for the $SU(4)$ infinite-$U$ Hubbard model is also well anticipated from former studies~\cite{Zhou2016mott,Zhou2018mott}, where people find VBS order at intermediate $U$, but the VBS order parameter decreases when people further increase $U$. Therefore, a transition from VBS to SL is expected, but whether it will happen at a finite larger $U$ or only happen in the infinite-$U$ limit is an open question. 

{\it Infinite-$U$ $SU(2N)$ extended Hubbard model on bipartite lattice}\,---\,
Further, we apply our projection approach to extended Hubbard models $H=H_t+H_U$, where the interaction part is defined as $H_U = U_\p\sum_\p (Q_\p - \nu_\p )^2 $, where $Q_\p$ is the plaquette charge operator defined on the elemental plaquette of the lattice, $Q_\p=\frac{1}{z}\sum_{i\in \p} n_i$, where the factor $z$ is used to normalize the filling of each plaquette as each site is shared by $z$ plaquettes; $z=4$ for square lattice and $z=3$ for honeycomb lattice. Similar to the Hubbard model, we have the following relation in the infinite-$U$ limit for the extended Hubbard model~\cite{suppl}:
\begin{equation}
 \left.e^{-\Delta_\tau U_\p (Q_{\p}- \nu)^2}\right|_{U_\p\rightarrow+\infty} 
=  \frac{1}{M}\sum_{s_\p=1}^M e^{\frac{\ii 2\pi z s_\p}{M} (Q_{\p}- \nu_\p)},
\end{equation}
with $M=\frac{z\eta\Nf}{2}+z|\tilde{\nu}_\p|+1$, where $\eta$ is the effective number of sites per plaquette, $\eta=1$ for square lattice and $\eta=2$ for honeycomb lattice. Here $\nu_\p$ is defined as the filling per plaquette, which is different from the $\nu$ defined in the onsite Hubbard model, where $\nu$ is the filling per site.
The cluster charge model is originally motivated to describe magic angle twisted bilayer graphene~\cite{Po2018origin,Xu2018kekule}, and we will explore the infinite-$U$ correlated ground state. Again, we consider several different kinds of fermiology, including $\square$-nesting-FS, $\square$-Dirac, and $\varhexagon$-Dirac. The possible ground states are listed in Table~\ref{table:tab2}. For $SU(2)$ and $SU(\ge 6)$, the ground states are the same with the Hubbard model as shown in Table~\ref{table:tab1}, while for the $SU(4)$ case with $\varhexagon$-Dirac dispersion, the extended Hubbard favors a Kekul\'e type VBS order. This is consistent with the large-$U$ result in Ref.~\cite{Liao2019valence}.

\begin{table}[t]
    \caption{Ground state of infinite-$U$ $SU(2N)$ extended Hubbard model [denoted as $SU(2N)$-$U_\p^\infty$] at half filling on bipartite lattice with different fermiology. For $N\ge 3$ we have VBS as the ground state for all three cases.} 
    \centering 
    \def\arraystretch{1.5}
    \begin{tabular}{c  c  c  c} 
    \hline\hline 
     $SU(2N)$-$U_\p^\infty$  & \ \  $\square$-nesting-FS \ \  & \ \ $\square$-Dirac \ \ & \ \ $\varhexagon$-Dirac \\ [0.0ex] 
    \hline 
     $SU(2)$   & \ \  N\'eel  &  \ N\'eel  & N\'eel  \\ [0.0ex] 
    \hline
     $SU(4)$  & \ \  N\'eel  &  \  SL?  & VBS \\ [0.0ex] 
    \hline
     $SU(\ge 6)$  & \ \  VBS  &  \  VBS  & VBS  \\ [0.0ex]
    \hline                    
    \end{tabular}
    \label{table:tab2} 
\end{table}

\begin{figure}[t]
\centering
\includegraphics[width=0.95\hsize]{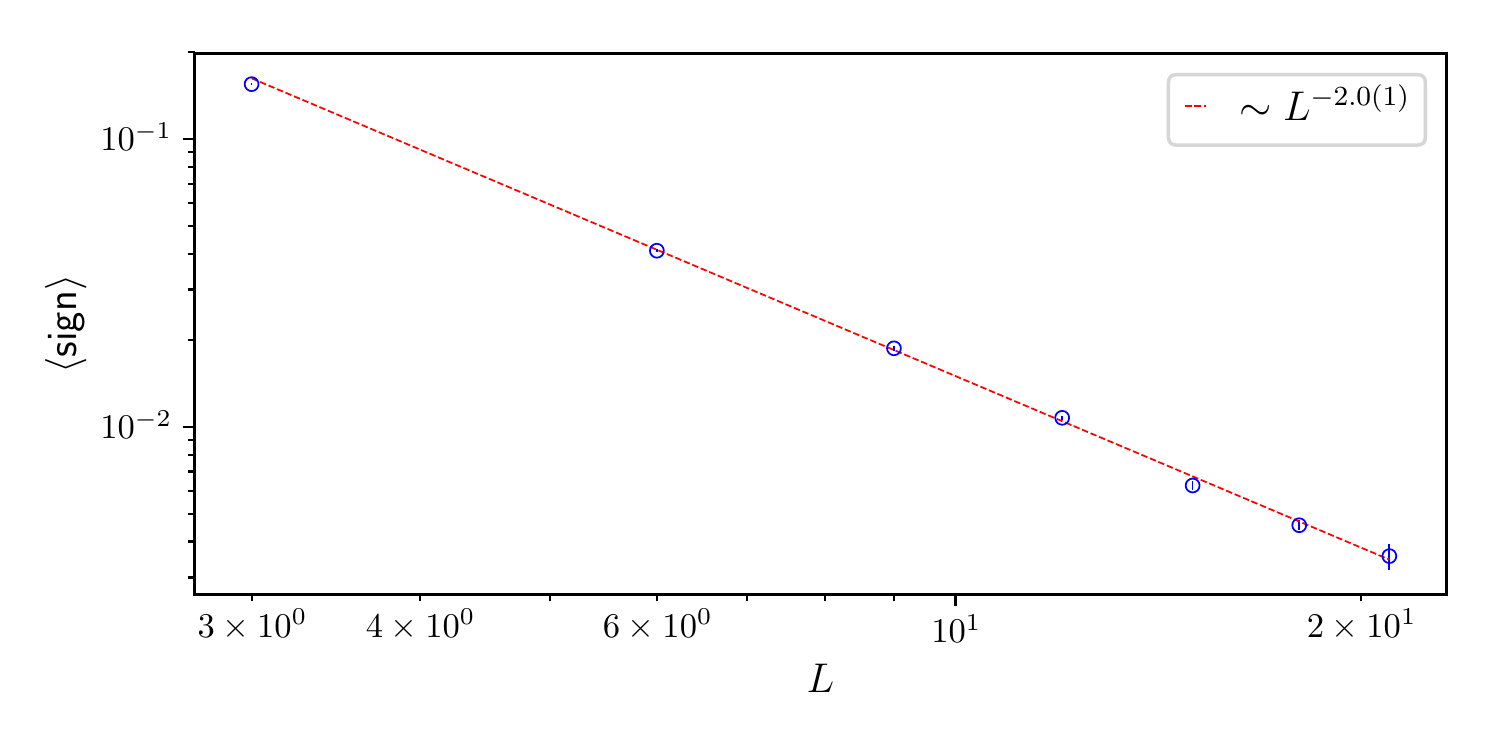}
\caption{Linear system size dependence of average sign $\langle \text{sign} \rangle$. The red dashed line is a power law fitting, getting a $L^{-2.0(1)}$ dependence. The power law system size dependence of average sign indicating an $\sim \frac{\ln(\beta N_s)}{\beta N_s}$ dependence of the free energy density difference of the original system and the bosonic reference system. }
\label{fig:fig4}
\end{figure}

{\it Strong coupling $SU(2N)$ extended Hubbard models with assisted hopping term on bipartite lattice}\,---\,
Another interesting application of our projection approach is for strong coupling $SU(2N)$ extended Hubbard models with assisted hopping term on bipartite lattice, where the kinetic part $H_t$ is turned off, $H \equiv H_U = U_\p\sum_\p (\tilde{Q}_\p - \nu_\p )^2 $ with $\tilde{Q}_\p \equiv Q_\p+\alpha T_\p$, and where the assisted hopping term $T_\p$ results from topological obstruction when people try to construct an effective real space lattice model for magic angle twisted bilayer graphene (TBG)~\cite{Kang2019strong} . We will focus on the $SU(4)$ case on a honeycomb lattice, which is directly related to magic angle TBG. It would be quite interesting to explore the possible ground states at each integer filling, where correlated insulator phases are found almost at all integer fillings ($\tilde{\nu}_\p=0,\pm 1,\pm 2, \pm 3$)~\cite{cao2018correlated,yankowitz2019tuning,lu2019superconductors}. We have the following relation to implement the projection~\cite{suppl}: 
\begin{equation}
  \left.e^{-\Delta_\tau U_\p (\tilde{Q}_{\p}- \nu_\p)^2}\right|_{U_\p\rightarrow+\infty} 
 = \frac{1}{M}\sum_{s_\p=1}^M e^{\frac{\ii 2\pi z s_\p}{M} (\tilde{Q}_{\p} - \nu_\p)}.
\end{equation}
As the kinetic part $H_t$ is turned off, we only have $\beta U_{\p}$ as an independent parameter and, in the DQMC simulation, we divide $\beta U_\p$ into $L_\tau$ slices $\beta U_\p \equiv L_\tau \Delta_\tau U_\p$, and we let $L_\tau$ scale with $L$, $L_\tau = 10L$. When we take the infinite-$U_{\p}$ limit, it corresponds to the zero temperature properties of the model. For $\tilde{\nu}_\p=0$, it is sign problem free, as pointed out by one of us~\cite{Liao2021correlation}. It favors an inter valley coherent state, when the kinetic part $H_t$ is added back, which breaks $SU(4)$ into two $SU(2)$ for each valley and a valley $U(1)$~\cite{Liao2021correlation}. For the strong coupling extended Hubbard model with assisted hopping term, we have full $SU(4)$, and we found an $SU(4)$ ferromagnetic state is stabilized for any finite $\alpha$. For other integer fillings, there is a sign problem. We define a reference \textit{bosonic} system with partition function $Z_b=\sum_c |w_c|$~\footnote{if $w_c$ is a complex number, we use $|\Re(w_c)|$.}, and the observables become $\langle O \rangle = \frac{\langle O \rangle_b}{\langle \text{sign} \rangle_b}$ where $\langle \cdots \rangle_b$ denotes sampling according to the reference \textit{bosonic} system, i.e., $\langle O \rangle_b = \sum_c O_c |w_c| /Z_b$. In general, $\langle \text{sign} \rangle_b$ decays exponentinally with system size, as $\langle \text{sign} \rangle_b \sim e^{-\beta N_s \Delta f}$~\cite{Troyer2005computational}, where $N_s$ is the total number of sites $\Delta f$ is the free energy density difference between the original system (with weight $w_c$) and the reference system (with weight $|w_c|$). However, we found a very interesting phenomenon at $|\tilde{\nu}_\p|=2$, i.e., the average sign decays algebraically instead of exponentially with system size as shown in Fig.~\ref{fig:fig4}, such that a power law computation complexity is expected and reliable QMC results can be obtained. Our simulation suggests an $SU(4)$ ferromagnetic state, confirming the analytical exact argument in Ref.~\cite{Kang2019strong}. 
We conjecture that the less severe sign problem may come from little fluctuations of the $SU(4)$ ferromagnetic ground state~\cite{Kang2019strong}. In detail, $\langle \text{sign} \rangle_b = \langle \text{sign}[\exp(4\ii\pi z\sum_{\p,l} s_{\p,l}/M)] \rangle_b$ and the sign problem is mild if $\sum_{\p,l} s_{\p,l}$ has little fluctuations due to an $SU(4)$ ferromagnetic ground state which usually has less fluctuations.
The algebraic sign behavior also implicitly indicates $\Delta f$ is $\sim \frac{\ln(\beta N_s)}{\beta N_s}$ small, such that a faithful mapping of a \textit{fermionic} system to a \textit{bosonic} system in the thermodynamic limit is obtained. 

{\it Discussion and conclusion}\,---\,
Our projection approach paves a way to study the infinite-$U$ Hubbard model at integer fillings. We also show how to apply our projection approach to the extended Hubbard model. As the infinite-$U$ Hubbard term is usually used to make a constraint on the local Hilbert space, such as for quantum spin models and Kondo lattice models, our projection approach can be further used to implement those constraints, such that they can be simulated in the framework of fermionic QMC simulations~\cite{Xu2021}.

One important issue of QMC simulations is the sign problem. It is generally believed that, if there is a sign problem, the average sign will decay exponentially with system size. Our finding provides a counterexample, inspiring a different direction to find Monte Carlo simulatable models, where the sign does not have to be always semipositive, as long as its average has an equal or better scaling than algebraic scaling with system size.

We further remark that our projection approach can also be extended to any rational filling, while the price is that we may need a high component of auxiliary fields depending on the filling factors~\cite{suppl}. In addition, we can also use a Hubbard-type term to impose the conservation of total number of particles, such that a hard constrained canonical ensemble Monte Carlo method is obtained, going beyond a soft constrained one~\cite{Wang2017finite}.

\begin{acknowledgments}
We acknowledge Y. Qi and T. Grover for stimulating discussions and F. Assaad for comments on the draft. X.Y.X. is sponsored by the Ministry of Science and Technology of China (Grant No. 2021YFA1401400), Shanghai Pujiang Program under Grant No. 21PJ1407200 and startup funds from SJTU. X.Y.X. also acknowledges support by the National Science Foundation under Grant No. DMR-1752417. This work used the Extreme Science and Engineering Discovery Environment (XSEDE)~\cite{xsede}, which is supported by National Science Foundation Grant No. ACI-1548562.
\end{acknowledgments}

\bibliographystyle{apsrev4-2}
\bibliography{main}

\newpage 
\clearpage
\onecolumngrid
\begin{center}
\textbf{Supplemental Material for ``Projection of Infinite-$U$ Hubbard Model and Algebraic Sign Structure"}
\end{center}
\setcounter{equation}{0}
\setcounter{figure}{0}
\setcounter{table}{0}
\setcounter{page}{1}
\makeatletter
\renewcommand{\thetable}{S\arabic{table}}
\renewcommand{\theequation}{S\arabic{equation}}
\renewcommand{\thefigure}{S\arabic{figure}}
\setcounter{secnumdepth}{3}

\section{Proof of Eq.(1), Eq.(2) and Eq.(3)}
Since the fermion occupation number of each site and flavor can only take values 0 or 1, it is easy to enumerate the values on each side of the identities, especially for Eq.(1) and Eq.(2). Further, note that the right hand side of those identities in Eq.(1), Eq.(2) and Eq(3) is a sum over geometric progression, which can be used to simplify the proof. For example, for Eq.(1), when $n_i = \nu$, both left hand side and right hand side are one; when $n_i \ne \nu$, the left hand side is zero, and the right hand side is
\begin{equation}
 \frac{1}{M}\sum_{s_i=1}^M e^{\frac{\ii 2\pi s_i}{M} (n_{i}- \nu)} = \frac{e^{\frac{\ii 2\pi }{M} (n_{i}- \nu)}\left( 1-e^{\ii 2\pi (n_{i}- \nu)}\right)}{1-e^{\frac{\ii 2\pi }{M} (n_{i}- \nu)}}
\end{equation}
which is always zero as long as: (1) $n_i - \nu$ is an integer; (2) the denominator never equals to zero when $n_i \ne \nu$. The first condition requires integer filling and the second condition can be easily satisfied by setting a large enough $M$ such that $\frac{|n_i - \nu|}{M}<1$ for any $n_i \ne \nu$. As $n_i$ can take values $0,1,\cdots,\Nf$, it is easy to check that the smallest $M$ we can take is $M=\frac{\Nf}{2}+|\nu-\frac{\Nf}{2}|+1$ for Eq.(1), and $M=\frac{z\eta\Nf}{2}+z|\nu_\p-\frac{\eta\Nf}{2}|+1$ for Eq.(2).

To get smallest $M$ for Eq.(3) is more involved, as it contains assistant hopping operator, which is defined as~\cite{Kang2019strong}
\begin{equation}
T_\p = \sum_{j\in \p, j=1}^6 \left( (-)^{j-1} e^{-\ii(-)^j \theta_\lambda}c_{j+1,\lambda,\sigma}^\dagger c_{j,\lambda,\sigma} + \text{h.c.} \right).
\end{equation}
The phase factor $e^{-\ii(-)^j \theta_\lambda}$ can be absorbed by a gauge transformation of fermion operator, therefore, there is a freedom in choosing $\theta_\lambda$ which is useful in the following. An interesting observation is that for any fermion occupation configurations, the assistant hopping part contributes zero value in total in the counting if we take $\theta_\lambda=\frac{\pi}{2}$. For example, considering following three configurations, one can easily check above statement, similarly for any other configurations. Therefore, $M$ can be set the same as in Eq.(2).
\begin{figure}[h]
\centering
\includegraphics[width=0.55\hsize]{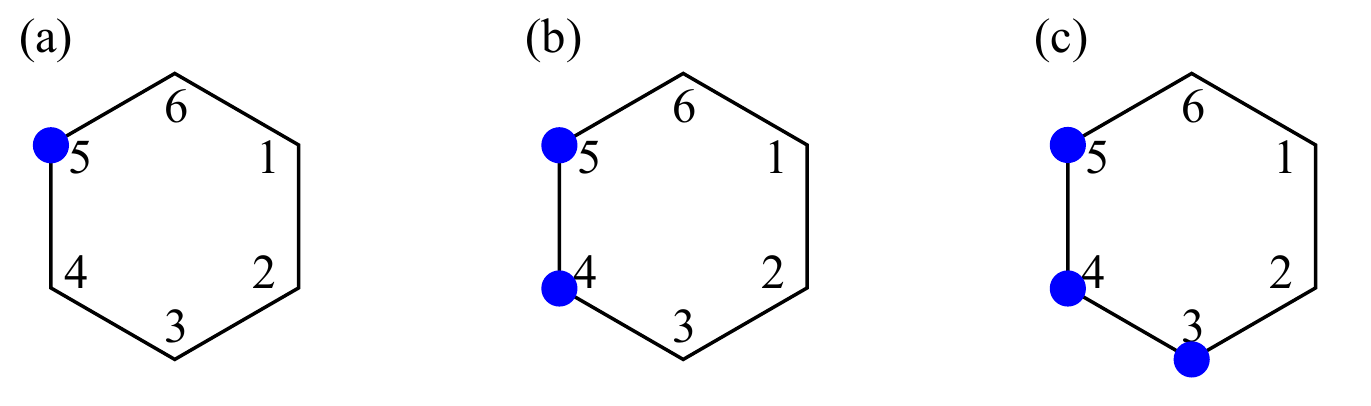}
\caption{Consider three typical fermion configurations for assistant hopping term. (a) Hopping $5\rightarrow 4$ gives $-e^{-\ii\theta_\lambda}$, hopping $5\rightarrow 6$ gives $e^{-\ii\theta_\lambda}$, so in total it gives zero. (b) Hopping $5\rightarrow 6$ gives $e^{-\ii\theta_\lambda}$, hopping $4\rightarrow 3$ gives $e^{\ii\theta_\lambda}$, so in total it gives zero for $\theta_\lambda = \frac{\pi}{2}$. (c) Hopping $5\rightarrow 6$ gives $e^{-\ii\theta_\lambda}$, hopping $3\rightarrow 2$ gives $-e^{-\ii\theta_\lambda}$, so in total it gives zero.}
\label{fig:fermionconf}
\end{figure}

\section{Generalize to rational filling}
The projection approach can be generalized to any rational filling, for example, for Eq.(1), when extend it to rational filling $\nu=\frac{p}{q}$, we have
\begin{equation}
\left.e^{-\frac{\Delta_\tau U}{2} (n_{i}- \frac{p}{q})^2}\right|_{U\rightarrow+\infty} =  \frac{1}{M}\sum_{s_i=1}^M e^{\frac{\ii 2\pi q s_i}{M} (n_{i}- \frac{p}{q})},
\end{equation}
with $M=\frac{q\Nf}{2}+q|\frac{p}{q}-\frac{\Nf}{2}|+1$.

\section{Zero temperature DQMC}
For zero temperature DQMC, the observables
\begin{equation}
\langle O \rangle = \frac{\langle \Psi_0 |O|\Psi_0 \rangle}{\langle \Psi_0 |\Psi_0 \rangle} = \lim_{\beta \rightarrow \infty } \frac{\langle \Psi_T |e^{-\frac{\beta}{2} H}Oe^{-\frac{\beta}{2} H}|\Psi_T \rangle}{\langle \Psi_T |e^{-\beta H}|\Psi_T \rangle}
\end{equation}
where $|\Psi_0\rangle$ is the ground state wavefunction, obtained by projection on a trial wave function $|\Psi_T\rangle$, $|\Psi_0\rangle=e^{-\frac{\beta}{2} H}|\Psi_T\rangle$. The projection time $\frac{\beta}{2}$ is set to be a large number to obtain the ground state.  After making similar Trotter decompositions, using same infinite-$U$ projection relation as in finite temperature case, and tracing out of fermion degrees of freedom, we have
\begin{equation}
\langle O \rangle = \frac{\sum_c O_c w_c}{\sum_c w_c}
\end{equation}
One can refer to Refs.~\cite{Sugiyama1986,Sorella1989a,Sorella1988numerical} for more details of zero temperature of DQMC. In our zero-temperature calculation, $\beta t = 60$ is large enough and  the trial wavefunction $|\Psi_T\rangle$ is set to be the ground state of non-interacting part. The trotter step is set as $\Delta_\tau t = 0.1$, and total number of time slices is set as $L_\tau = \beta /\Delta_\tau = 600$.

\section{Softening of static form factor in possible SL phase}
The static form factor for spin or bond ($C_{N/B}(\vec{q})$) are defined as
\begin{equation}
C_{N/B}(\vec{q}) = \frac{1}{L^4} \sum_{i,j} C_{N/B}(\vec{r}_i-\vec{r}_j)e^{-\ii \vec{q}\cdot(\vec{r}_i-\vec{r}_j)}
\end{equation}
For $SU(\Nf)$ Hubbard model with $\square$-Dirac and $\varhexagon$-Dirac fermiology, there is sharp peak in spin static factor at $\Nf=2$ and sharp peak in bond static factor at $\Nf=6$. The static form factor of spin and bond at $\Nf=4$ shows some softening, indicating absence of any possible trend to spin or bond orders.
\begin{figure}[h]
\centering
\includegraphics[width=0.55\hsize]{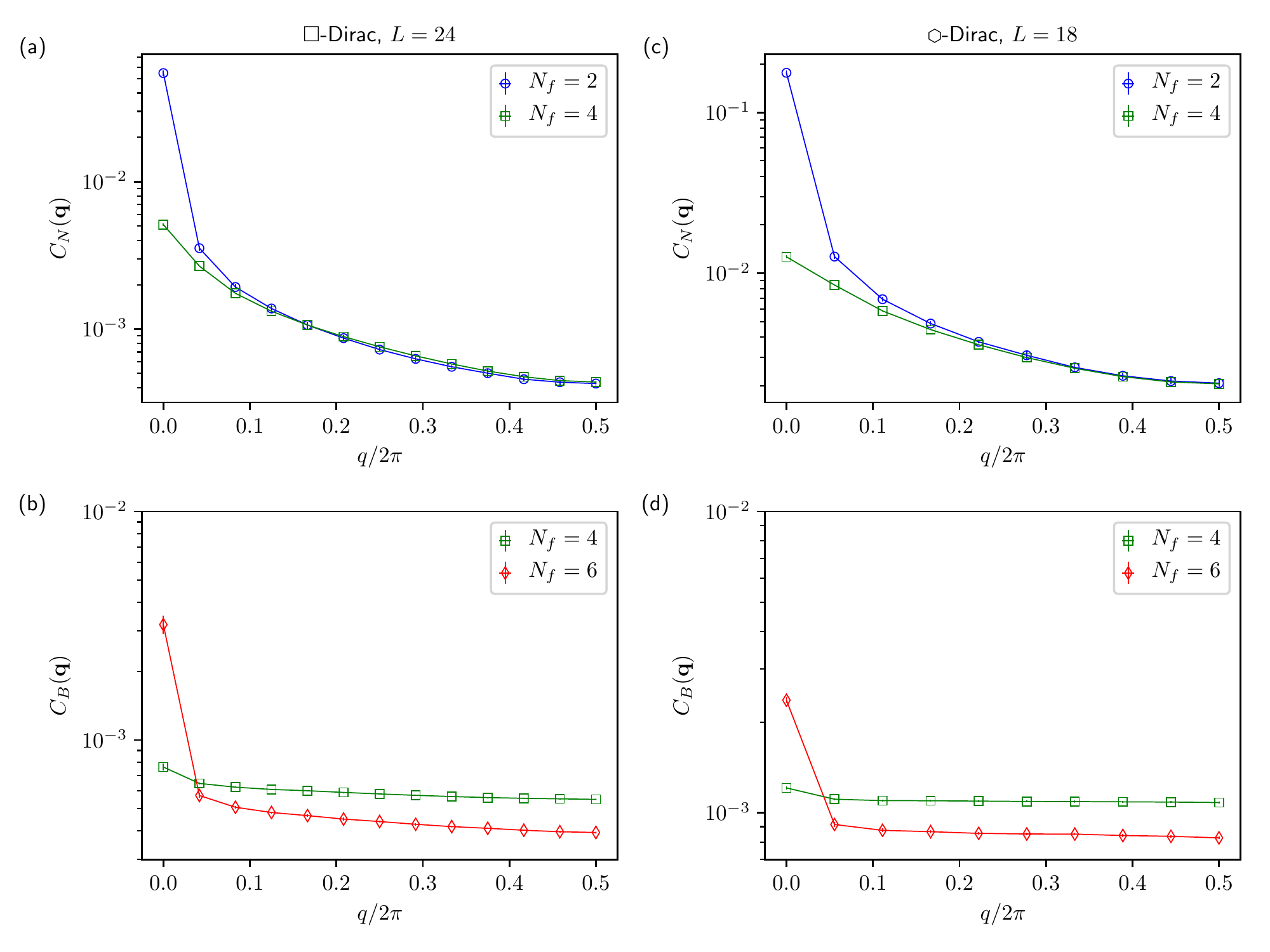}
\caption{Static form factor along $q_x$ direction. (a) Spin form factor for $\square$-Dirac, $L=24$. (b) Bond form factor for $\square$-Dirac, $L=24$. (c) Spin form factor for $\varhexagon$-Dirac, $L=18$. (d) Bond form factor for $\varhexagon$-Dirac, $L=18$.}
\label{fig:figkpath}
\end{figure}

\end{document}